\documentclass[aps,twocolumn,showpacs,preprintnumbers,prl,superscriptaddress,groupedaddress,10pt]{revtex4-1}
\usepackage[utf8]{inputenc}
\usepackage{graphicx,amssymb,amsmath,amsthm,amsfonts,epsfig,times,natbib}

\usepackage[usenames,dvipsnames]{color}
\usepackage{epstopdf}
\usepackage{amsmath,amssymb}
\usepackage{tensor}
\usepackage{mathtools}
\usepackage{amsbsy}
\usepackage{bm,url}
\usepackage[linktocpage]{hyperref}

\usepackage[scaled]{beramono}
\usepackage[T1]{fontenc}

\definecolor{oxfordblue}{rgb}{0.0, 0.13, 0.28}
\definecolor{burgundy}{rgb}{0.5, 0.0, 0.13}
\definecolor{darkolivegreen}{rgb}{0.33, 0.42, 0.18}
\definecolor{darkblue}{rgb}{0,0,0.5}
\definecolor{richcarmine}{rgb}{0.84, 0.0, 0.25}
\definecolor{darkblue}{rgb}{0,0,0.5}
\definecolor{venetianred}{rgb}{0.78, 0.03, 0.08}
\definecolor{skobeloff}{rgb}{0.0, 0.48, 0.45}
\hypersetup{colorlinks=true, citecolor=darkblue, linkcolor=darkblue,
urlcolor = darkblue}

\newcommand{\ben}{\begin{enumerate}}
\newcommand{\een}{\end{enumerate}}

\def\be{\begin{equation}}
\def\ee{\end{equation}}
\def\bea{\begin{eqnarray}}
\def\eea{\end{eqnarray}}

\newcommand{\beq}{\begin{eqnarray}}
\newcommand{\eeq}{\end{eqnarray}}
\newcommand{\ba}{\begin{align}}
\newcommand{\ea}{\end{align}}

\begin{document}

\title{Multipolar boson stars: macroscopic Bose-Einstein condensates akin to hydrogen orbitals}

\author{
C. A. R. Herdeiro$^{1}$,
J. Kunz$^{2}$,
I. Perapechka$^{3}$,
E. Radu$^{1}$ and
Ya. Shnir$^{4}$
}

\affiliation{${^1}$ Departamento de Matem\'atica da Universidade de Aveiro and CIDMA,
Campus de Santiago, 3810-183 Aveiro, Portugal}
\affiliation{${^2}$ Institute of Physics, University of Oldenburg, Oldenburg, 26111, Germany }
\affiliation{${^3}$ Department of Theoretical Physics and Astrophysics, Belarusian State University }
\affiliation{${^4}$ BLTP, JINR, Joliot-Curie 6, Dubna 141980, Moscow Region, Russia }


\date{August 2020}

\begin{abstract}
Boson stars are often described as macroscopic Bose-Einstein condensates. By accommodating large numbers of bosons in the same quantum state, they materialize macroscopically the  intangible probability density cloud of a single particle in the quantum world. We take this interpretation of boson stars one step further. We show, by explicitly constructing  the fully non-linear solutions, that static (in terms of their spacetime metric, $g_{\mu\nu}$) boson stars, composed of a single complex scalar field, $\Phi$, can have a non-trivial multipolar structure, yielding the same morphologies  for their energy density as those that elementary hydrogen atomic orbitals have for their probability density. This provides a close analogy between the elementary solutions of the non-linear Einstein--Klein-Gordon theory, denoted $\Phi_{(N,\ell,m)}$, which could  be realized in the macrocosmos, and those of the linear Schr\"odinger equation in a Coulomb potential, denoted $\Psi_{(N,\ell,m)}$, that  describe the microcosmos. In both cases, the solutions are classified by a triplet of quantum numbers $(N,\ell,m)$. In the gravitational theory, multipolar boson stars can be interpreted as individual bosonic lumps in equilibrium; remarkably, the (generic) solutions with $m\neq 0$ describe gravitating solitons $[g_{\mu\nu},\Phi_{(N,\ell,m)}]$ without \textit{any continuous symmetries}. Multipolar boson stars analogue to hybrid orbitals are also constructed.
\end{abstract}




\maketitle
{\bf Introduction.}
Atomic orbitals are solutions of the linear, non-relativistic Schr\"odinger equation in an appropriate electromagnetic potential. Their morphologies have become iconic images shaping scientific insight and popular visualization about the microscopic world, see $e.g.$~\cite{Thaller,orbitals}. Yet, they are  intangible probability clouds. The hydrogen atom, in particular, provides the cornerstone orbitals, $\Psi_{(N,\ell,m)}$. The properties of $\Psi_{(N,\ell,m)}$ are closely connected to the separability of the wave function into a spherical harmonic and a radial function, where the nodal structure is defined by  the standard quantum numbers $(N,\ell,m)$~\footnote{In this discussion we ignore quantum spin.}. At a deeper level, the $\Psi_{(N,\ell,m)}$ properties result from the explicit and hidden symmetries (yielding an $SO(4)$ group) provided by the Coulomb potential~\cite{jones}.

In the macroscopic realm, on the other hand, the relativistic generalization of the Schr\"odinger equation, $i.e.$ the Klein-Gordon equation, when minimally coupled to Einstein's gravity has produced the most paradigmatic example of a self-gravitating soliton: boson stars (BSs)~\cite{Kaup:1968zz,Ruffini:1969qy,Schunck:2003kk,Liebling:2012fv}. Suggested as dark matter lumps, if ultralight bosons exist~\cite{Li:2013nal,Suarez:2013iw,Hui:2016ltb}, their  bosonic character allows the individual quanta to inhabite the same state. BSs are envisaged as a macroscopic Bose-Einstein condensates, materializing as a macroscopic energy distribution the intangible concept of a quantum probability density.

In  this letter we lay down a foundational construction to corroborate the interpretation that BSs are macroscopic atoms. In the Einstein--Klein-Gordon model, with a single, complex, massive, free scalar field $\Phi$, the only static BSs~\footnote{Static in  the sense their geometry admits an everywhere timelike, hypersurface orthogonal, Killing vector field. But $\Phi$ is not invariant under this Killing vector field.} known so far are spherically  symmetric.  Thus, they are macroscopic $Ns$-orbitals, where  $N-1$ describes the number of radial nodes~\footnote{The discussion of radial nodes in non-spherical symmetry is addressed below.}. We shall show, by explicit construction, that BSs corresponding to all $(N,\ell,m)$ hydrogen-orbitals exist, with identical morphologies to their microscopic counterparts, in spite of the very different mathematical structure of both models. These \textit{multipolar BSs} shall be denoted as $\Phi_{(N,\ell,m)}$.

{\bf The framework.}
The simplest BSs, often called \textit{mini-BSs}~\cite{Kaup:1968zz,Ruffini:1969qy}, are solutions of the Einstein--(complex, massive)Klein-Gordon model, described by the action $\mathcal{S}=\int  d^4x \sqrt{-g}\mathcal{L}$. The Lagrangian  density is:
\begin{equation}
\label{action}
\mathcal{L}= \frac{R}{16\pi G}
   -\frac{1}{2} g^{\alpha\beta}\left(\partial_\alpha \Phi^* \partial_\beta\Phi + \partial_\beta\Phi^* \partial_\alpha\Phi \right) - \mu^2 \Phi^*\Phi
\  ,
\end{equation}
where $R$ is the Ricci scalar of $g_{\alpha\beta}$, $G$ is Newton's constant, $\mu$ is the scalar field mass and $^*$  denotes complex conjugation.

This action is invariant under the global $U(1)$ transformation $\Phi\rightarrow e^{i\chi}\Phi$, where $\chi$ is constant, yielding a conserved 4-current, $D_\alpha j^\alpha=0$, where $j^\alpha\equiv-i (\Phi^* \partial^\alpha \Phi-\Phi \partial^\alpha \Phi^*)$.
The associated conserved quantity, obtained by integrating the timelike component of this 4-current in a spacelike slice $\Sigma$ is the \textit{Noether charge}, $Q=\int_{\Sigma}~j^t$. At a microscopic level, this Noether charge counts the number of scalar particles, a relation that can be made explicit by quantizing the scalar field.

The key feature of BSs is the existence of a harmonic time dependence for $\Phi$, analogue to that of stationary states in quantum mechanics (QM). Consider a background geometry written in spherical-like coordinates $(t,r,\theta,\varphi)$ with their usual meanings. For BSs the scalar field has the form
\be
\Phi=e^{-i\omega t}f(r,\theta,\varphi) \ ,
\label{phi}
\ee
where $\omega$ is the oscillation frequency of the star and $f(r,\theta,\varphi)$ a real spatial profile function. The oscillation, however, occurs only for the field, in the same way the probability \textit{amplitude}  of stationary states oscillates in QM. The physical quantities of the BSs, such as its energy-momentum tensor, are time-independent, like the probability \textit{density} in QM. The oscillating $\Phi$-amplitude is crucial to circumvent Derrick-type~\cite{Derrick:1964ww} virial identities; it allows the existence of stationary solitons of \eqref{action}.

{\bf The spherical sector: $\Phi_{(N,0,0)}$.}
The known static BSs are spherically symmetric. There is a countable infinite number of families of such BSs, labelled by the number of radial nodes $n\in \mathbb{N}_0$. For sharpening  the parallelism with QM we introduce an  integer $N\in \mathbb{N}$, which for spherical BSs is $N=n+1$. The fundamental family has $N=1$ (no nodes); excited families have $N>1$. For each family, solutions exist for an interval of frequencies $\omega\in [\omega_{\rm min}^{(N,0,0)},\mu]$. For instance, the fundamental family has $\omega_{\rm min}^{(1,0,0)}=0.768\mu$~\cite{Herdeiro:2017fhv}. The upper limit is universal: $\omega\leqslant \mu$ is a bound state condition, as it is clear from the asymptotic decay of the field: $\Phi\sim e^{-r\sqrt{\mu^2-\omega^2}}/r$. The $\Phi_{(N,0,0)}$ BS family corresponds to the $Ns$-orbital in hydrogen. Observe that for the latter, however, the frequency is a unique number, determined by $N$. By contrast, the $\Phi_{(N,0,0)}$ BSs are, for each $N$, a continuous family, existing for an interval of frequencies. This is likely a consequence of the non-linear structure of the BS model. This distinction between the single frequency $\Psi_{(N,0,0)}$ and the multi-frequency $\Phi_{(N,0,0)}$ will remain when introducing non-trivial $(\ell,m)$.

{\bf The non-spherical sector: introducing $(\ell,m)$.}
Scalar multipoles are typically described by spherical harmonics. The $real$ spherical harmonics
 $Y_{\ell m}(\theta,\varphi)$ are proportional to $P_\ell^{m}(\cos \theta)\cos m\varphi$,
where $P_\ell^{m}$ are the associated Legendre polynomials
and $\theta,\varphi$ are the usual angles on the $S^2$; $\ell ,m$ are integers with $\ell \geqslant m$ and we take $m\geqslant 0$
without any loss of generality.

For odd-$\ell$, $Y_{\ell m}$ are parity-odd and vanish at $\theta=\pi/2$;
for even-$\ell$ $Y_{\ell m}$ are parity-even and $\mathbb{Z}_2$ symmetric under a reflection along the
equatorial plane.
For any $\ell,m$,
$Y_{\ell m}$  has $2m$ $\varphi$-zeros, each describing a nodal longitude line and
$\ell - m$ $\theta$-zeros, each yielding a nodal latitude line.
These nodal distributions define an $(\ell,m)$ mode in the non-linear theory, where $\Phi$ is, in general, no longer described by a single $Y_{\ell m}$.

{\bf Gravity-regularization.}
Consider, for the moment,
the Klein-Gordon equation derived from (\ref{action}), $\Box\Phi=\mu^2\Phi$, on Minkowski spacetime  in spherical coordinates, taking $\Phi$ as a test field.
The  $Y_{\ell m}$  are a complete basis on $S^2$. Thus, for any given $\omega$, the general solution of this linear equation is of the form \eqref{phi} with
$f=  \sum_{\ell,m }R_{\ell}(r) Y_{\ell m}(\theta,\varphi)$.  The regular at $r\to \infty$ (real) radial amplitude is
\be
\label{exp2}
  R_{\ell}(r) =
 \frac{c}{\sqrt{r}}
 K_{\frac{1}{2}+\ell}(r\sqrt{\mu^2-\omega^2}) \ ,
\ee
where $c$ is an arbitrary constant and $K_{\ell+\frac12}(r)$ is the modified Bessel function of the first kind of order $\ell$.
This amplitude diverges  at the origin $r=0$; however, the
backreaction in Einstein's gravity regularizes the origin singularity. As such, the
spherical BSs can be viewed as the non-linear (regular) realization of the linear
(irregular) $K_{1/2}$ with $\ell=0=m$. As it turns out,
the gravitational backreaction can regularize the origin
singularity for \textit{any} $(\ell,m)$-harmonic.
This leads to multipolar BSs.
 Considering the backreaction, however, the scalar model becomes non-linear. Thus, the angular dependence of the resulting scalar field is no longer that of a pure
$Y_{\ell m}$ harmonic; it becomes a superposition of harmonics.
The construction reveals, nonetheless, that the gravitating solutions preserve the discrete
symmetries and nodal structure of the original $Y_{\ell m}$. For $m=0$, these configurations are axially symmetric.
In the generic  $m \neq 0$ case, only discrete symmetries remain.

{\bf Constructing multipolar BSs: ansatz and approach.}
Allowing an angular dependence for the BSs requires considering a metric ansatz with sufficient generality. In particular we do not assume any spatial isometries.  In the absence of analytic methods to tackle the fully non-linear Einstein-Klein-Gordon
solutions in the absence of symmetries, we shall resort to numerical methods~\footnote{Even spherically symmetric BSs are only known numerically.} - see also the construction in~\cite{Yoshida:1997jq}.

We consider a metric ansatz
with seven $(r,\theta,\varphi)$-dependent functions, $F_1,F_2,F_3,F_0,S_1,S_2,S_3$~\footnote{This ansatz is compatible with an energy momentum with $T^t_\varphi=T^t_\theta=T^t_r=0$; thus the solutions carry no momentum.}:
\begin{eqnarray}
\label{metric}
&&ds^2=
-F_0(r,\theta,\varphi)  dt^2
\\
\nonumber
&&
+ F_1(r,\theta,\varphi)dr^2+F_2(r,\theta,\varphi)\left[r d\theta+S_1(r,\theta,\varphi) dr \right]^2
\\
\nonumber
&&
+F_3(r,\theta,\varphi)  \big[r \sin \theta d\varphi+S_2(r,\theta,\varphi) dr+S_3(r,\theta,\varphi) r d\theta \big]^2
\ ,
\end{eqnarray}
together with the scalar field ansatz~\eqref{phi}. The resulting intricate set of partial differential equations (PDEs) is tackled by employing the Einstein-De Turck approach~\cite{Headrick:2009pv,Adam:2011dn}, in which the Einstein equations obtained from~\eqref{action} are replaced by
\begin{eqnarray}
\label{EDT}
R_{\mu\nu}-\nabla_{(\mu}\xi_{\nu)}= 8 \pi G \left(T_{\mu\nu}-\frac{1}{2}T  g_{\mu\nu}\right) \ .
\end{eqnarray}
  $\xi^\mu$ is  defined as
$
\xi^\mu\equiv g^{\nu\rho}(\Gamma_{\nu\rho}^\mu-\bar \Gamma_{\nu\rho}^\mu),
$
where
$\Gamma_{\nu\rho}^\mu$ is the Levi-Civita connection associated to the
spacetime metric $g$ that one wants to determine.
Also,  $\bar g$ is a reference metric  (with  connection $\bar \Gamma_{\nu\rho}^\mu$), which,
for the solutions in this work is the Minkowski line element.
Solutions to (\ref{EDT}) solve the Einstein equations
iff $\xi^\mu \equiv 0$ everywhere on the manifold.

Multipolar BSs with $m=0$ will be axi-symmetric. They can be studied within this framework by setting $S_2=S_3=0$ and taking
all other functions to depend on $(r,\theta)$ only.

{\bf Boundary conditions (BCs) and numerics.}
With the described setup, the problem reduces to
solving a set of eight PDEs with suitable BCs.
The latter are found from an approximate solution on the
boundary of the domain of integration compatible with $\xi^\mu = 0 $, regularity and asymptotic flatness.

We consider solutions with $m\geqslant 0$, a scalar field with $(-1)^\ell$ parity under reflections along the equatorial plane
($\theta=\pi/2$)
and two $\mathbb{Z}_2$-symmetries $w.r.t.$ the $\varphi-$coordinate.
Then the domain of integration  for the $(\theta,\varphi)$-coordinates
is $[0,\pi/2]\times [0,\pi/2]$, while $0\leqslant r<\infty$. The BCs are as follows: $(i)$ at $r=\infty$,  $F_1= F_2=  F_3= F_0=1$,
$S_1=S_2= S_3=0$ and $\phi=0$; $(ii)$ at $r=0$, $\partial_r F_1= \partial_r F_2=  \partial_r F_3=  \partial_r F_0=0$, $\partial_r S_1=  \partial_r S_2=  \partial_r S_3=0$ and $\phi=0$, except for axially symmetric odd-chains (as described below), for which $\partial_r \phi=0$; $(iii)$ at  $\theta=0$, $\partial_\theta F_1= \partial_\theta F_2=  \partial_\theta F_3= \partial_\theta F_0=0$,
$S_1=S_2= \partial_\theta S_3=0$ and $\phi=0$,
 except if $\ell+m$ is an even number,
in which case we impose $\partial_\theta \phi=0$;
$(iv)$ at $\theta=\pi/2$, $\partial_\theta F_1= \partial_\theta F_2=  \partial_\theta F_3= \partial_\theta F_0=0$,
$S_1=\partial_\theta  S_2=   S_3=0$ and $\partial_\theta  \phi=0$; $(v)$ at $\varphi=0$, $\partial_\varphi F_1= \partial_\varphi F_2=  \partial_\varphi F_3= \partial_\varphi F_0=0$,
$\partial_\varphi S_1=   S_2=   S_3=0$ and $\partial_\varphi \phi=0$; $(vi)$ at  $\varphi=\pi/2$, $\partial_\varphi F_1= \partial_\varphi F_2=  \partial_\varphi F_3= \partial_\varphi F_0=0$, $\partial_\varphi S_1=   S_2=   S_3=0$ and either  $ \phi=0$ for odd $m$, or  $\partial_\varphi \phi=0$ for even $m$.

The field equations are discretized on a $(r, \theta,\varphi)$ grid with
$N_r\times N_\theta \times N_\varphi$
points.
The grid spacing in the $r$-direction is non-uniform, whilst the values of the grid points in the angular
directions are uniform.
For the 3D problem, typical grids have sizes
$\sim 100 \times 30 \times 30$.
The resulting system is solved
iteratively using the Newton-Raphson
 method  until convergence is achieved.
The professional PDE solver \textsc{cadsol}~\cite{schoen} and
  the Intel \textsc{mkl pardiso}~\cite{pardiso} sparse direct solvers were both emplyed in this work.
For the solutions herein,
 the typical numerical error is estimated to be
$\lesssim 10^{-3}$.

Natural units, set by $\mu$ and $G$, are used. Dimensionless variables, $e.g.$,  $r\to r/\mu,~~
\phi \to \phi /\sqrt{4\pi G},~~
\omega \to \omega/\mu$ are employed in the numerics.
 As a result, all physical quantities of interest are
 expressed in units set by $\mu$ and $G$. The only input parameter is the (scaled) frequency, $\omega$.

As for spherical BSs, the multipolar BSs possess two global ``charges": the ADM mass $M$ and the Noether charge $Q$.
The ADM mass
is either read off from the far field asymptotics $-g_{tt}=-F_0 = 1-2M/r+\dots$, or computed as the volume integral $
M=-\int_{\Sigma} dS_\alpha (2T_{\beta}^\alpha \xi^\beta-T\xi^\alpha)$, where $\xi=\partial/\partial t$ is the everywhere timelike Killing vector field.

\begin{figure}[b!]
\begin{center}
\includegraphics[width=0.45\textwidth]{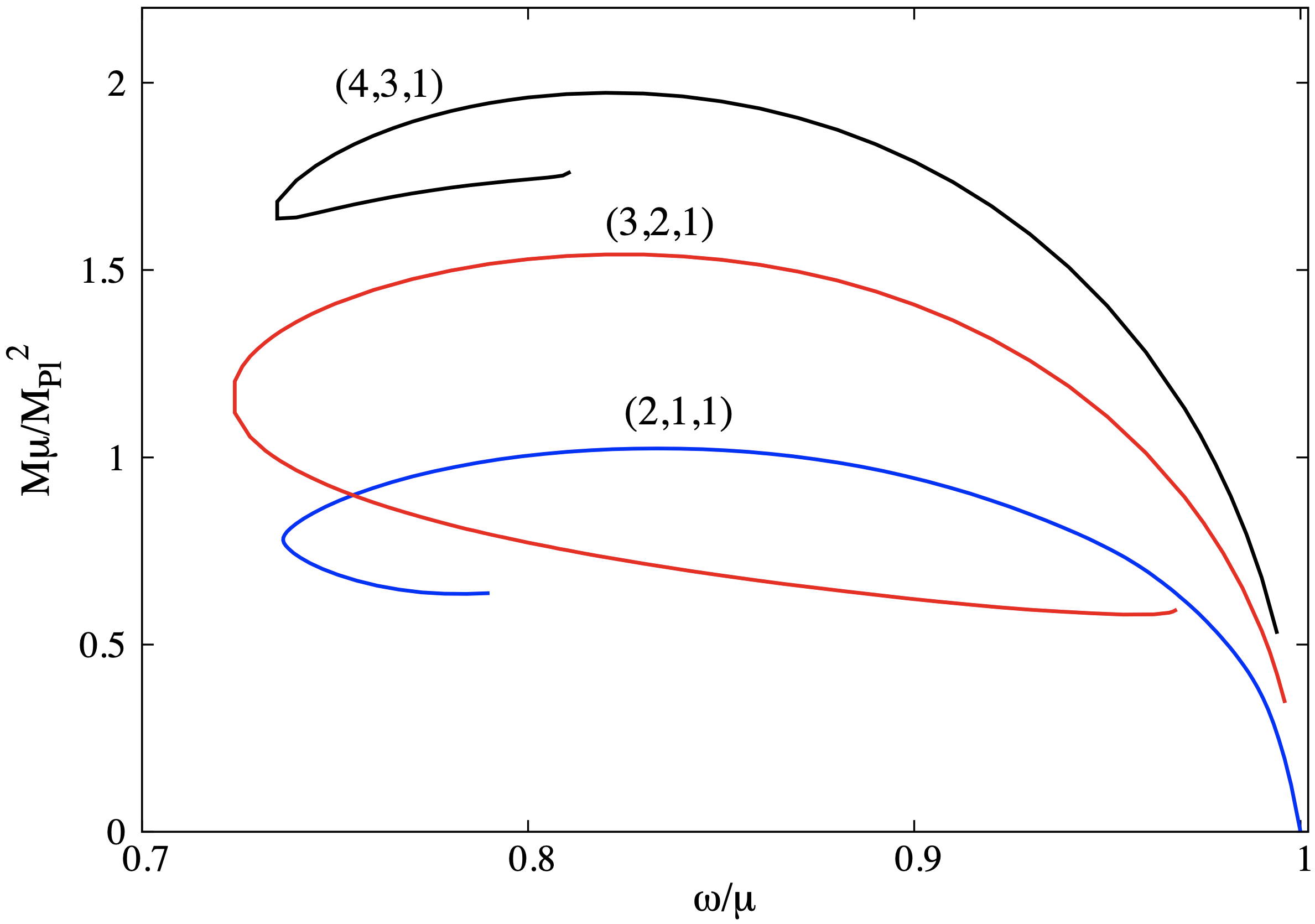}
\caption{\small Domain of existence of $\Phi_{(\ell+1,\ell,1)}$ multipolar BSs.
The configurations with $\ell= 1$ have a $U(1)$ isometry while those
with $\ell= 2,3$ possess  discrete symmetries only. }
\label{fig1}
\end{center}
\end{figure}

{\bf Multipolar BSs: Domain of existence and morphology.}
We have studied in detail the families of solutions of $\Phi_{(N,\ell,m)}$
BSs for a variety of $(N,\ell,m)$ values. $(\ell,m)$ are defined from the single $Y_{\ell m}$ present in the (irregular) flat spacetime limit. $N$ is defined by assigning the number of nodes along the half $z$-axis, $z\in ]0,\infty[$ to be $N-\ell-1$~\footnote{In the hydrogen atom this is the number of radial nodes.}.
In all $\Phi_{(N,\ell,m)}$  studied, the domain of existence spans a range of frequencies, yielding a spiraling curve in a $M$ $vs.$ $\omega$ diagram. In Fig.~\ref{fig1}
this is illustrated for $N=\ell+1$ (solutions without radial nodes), $m=1$ and for $\ell= 1,2,3$.
Qualitatively similar curves are found for all $\Phi_{(N,\ell,m)}$ BSs,
albeit secondary branches (obtained after each backbending of the curve, when an extremum of $\omega$ is reached)
are more difficult to explore; so only up to the second branch is shown in  Fig.~\ref{fig1}. Plotting  the  $Q$  (instead of $M$) also yields similar curves.

Fig.~\ref{fig1} shows that, as for $\Phi_{(N,0,0)}$ BSs, the maximum frequency is universal, but the minimum frequency, $\omega_{\rm min}^{(N,\ell,m)}$,  is $(N,\ell,m)$ dependent. It also illustrates the general trend that increasing any of the quantum numbers, the maximal mass of the family increases.

\begin{figure}[h!]
\begin{center}
\includegraphics[width=0.45\textwidth]{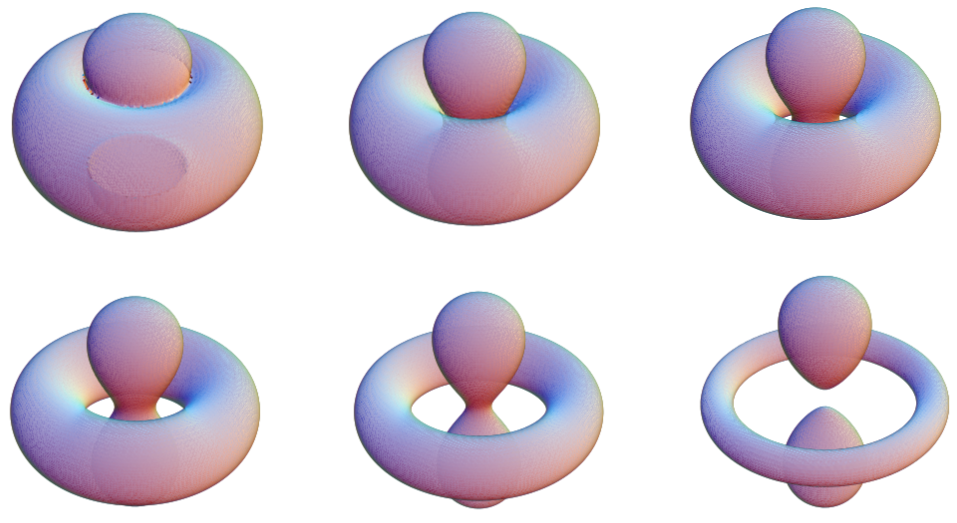}
\caption{\small Surfaces of constant energy density $\epsilon$ for a $\Phi_{(3,2,0)}$ BS, with $\omega=0.9$. $\epsilon$ increases, from left to right, top to bottom.}
\label{fig2}
\end{center}
\end{figure}

Along any fixed $\Phi_{(N,\ell,m)}$ family the BSs vary in size (in the scale fixed by $\mu$),
 but their morphology remains unchanged.
To analyse this morphology we  examine $e.g.$ the surfaces of constant energy density $\epsilon\equiv -T_t^t$;
but the same result is found when considering instead the Noether charge density.
In Fig.~\ref{fig2} we exhibit several surfaces of constant $\epsilon$ for $\Phi_{(3,2,0)}$.
Increasing the energy density clearly distinguishes several individual lumps.
This is a general feature: multipolar BSs have a well defined multicomponent
structure in the regions of larger energy density.

\begin{widetext}

\begin{figure}[t!]
\begin{center}
\includegraphics[width=0.95\textwidth]{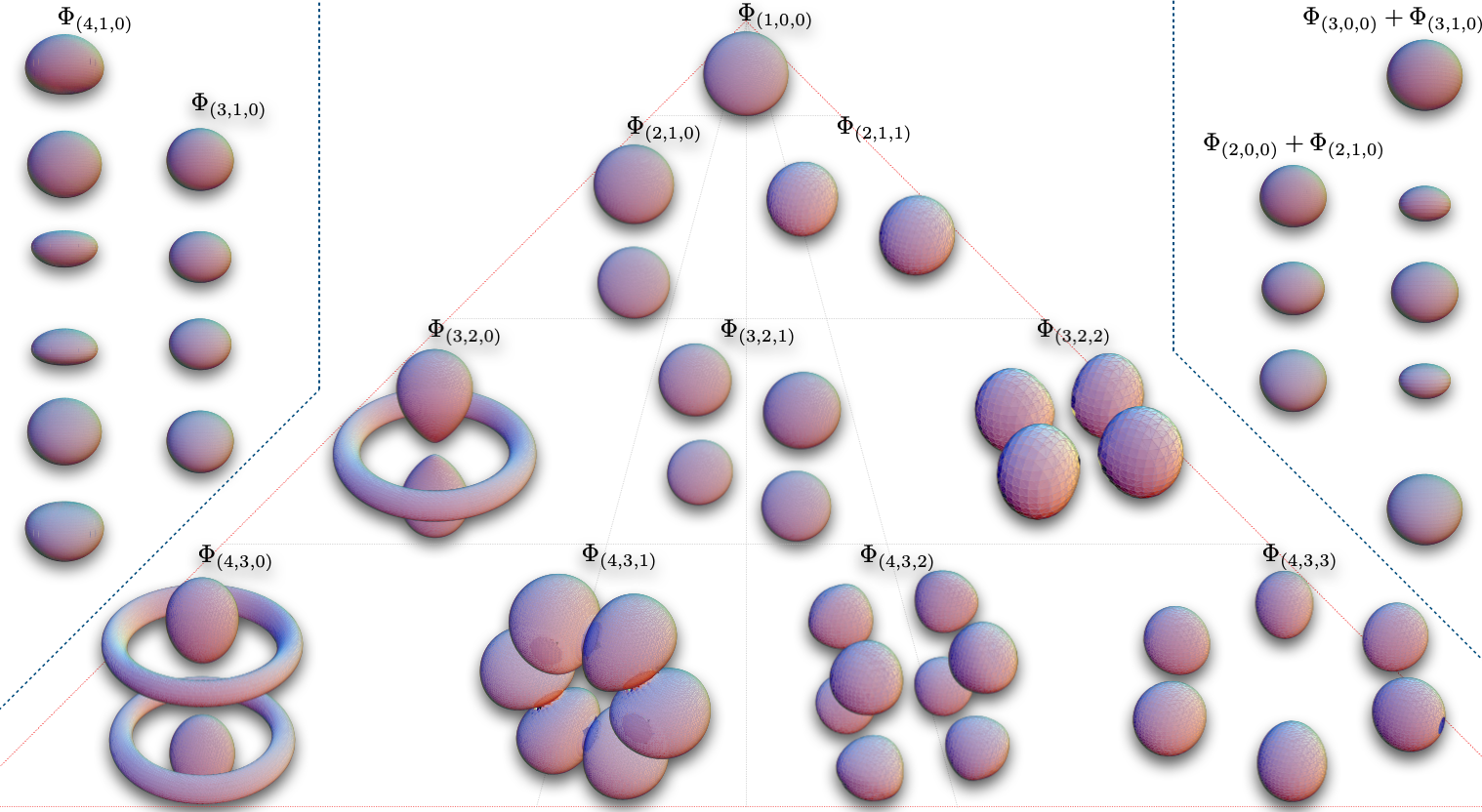}
\caption{\small Surfaces of constant energy density for a selection of $\Phi_{(N,\ell,m)}$. The  hydrogen orbitals-like morphology is unmistakable, see $e.g.$~\cite{orbitals}.}
\label{fig3}
\end{center}
\end{figure}

\end{widetext}

In Fig.~\ref{fig3} we provide an overview of a selection of multipolar BSs. The figure unveils an uncanning similarity with hydrogen orbits - see $e.g.$~\cite{orbitals}.

The multipolar BSs in the central triangle of Fig.~\ref{fig3} are of the form $\Phi_{(\ell+1,\ell,m)}$,
with $0\leqslant m\leqslant \ell$.
These have $N-\ell-1=0$ and, in this sense, they do not have "radial" nodes.
 $\Phi_{(1,0,0)}$ corresponds to the 1$s$-orbital, $\Phi_{(2,1,0)}$ to the 2$p$-orbital.
Alternatively,  $\Phi_{(1,0,0)}$ is the monopolar BS,  $\Phi_{(2,1,0)}$ is a dipole BS, and so on.
Generically, for  $m \neq 0$, only discrete symmetries exist (with some known  exceptions,
$e.g.$ $\Phi_{(2,1,1)}$, which is a rotation of $\Phi_{(2,1,0)}$ and thus possesses a $U(1)$ isometry).
 In  each case,  non-linearity  implies that the angular distribution is a superposition of harmonics; nonetheless, the corresponding $(\ell,m)$-mode shapes the  morphology of the BS in the larger energy density regions.

\begin{figure}[h!]
\begin{center}
\includegraphics[width=0.45\textwidth]{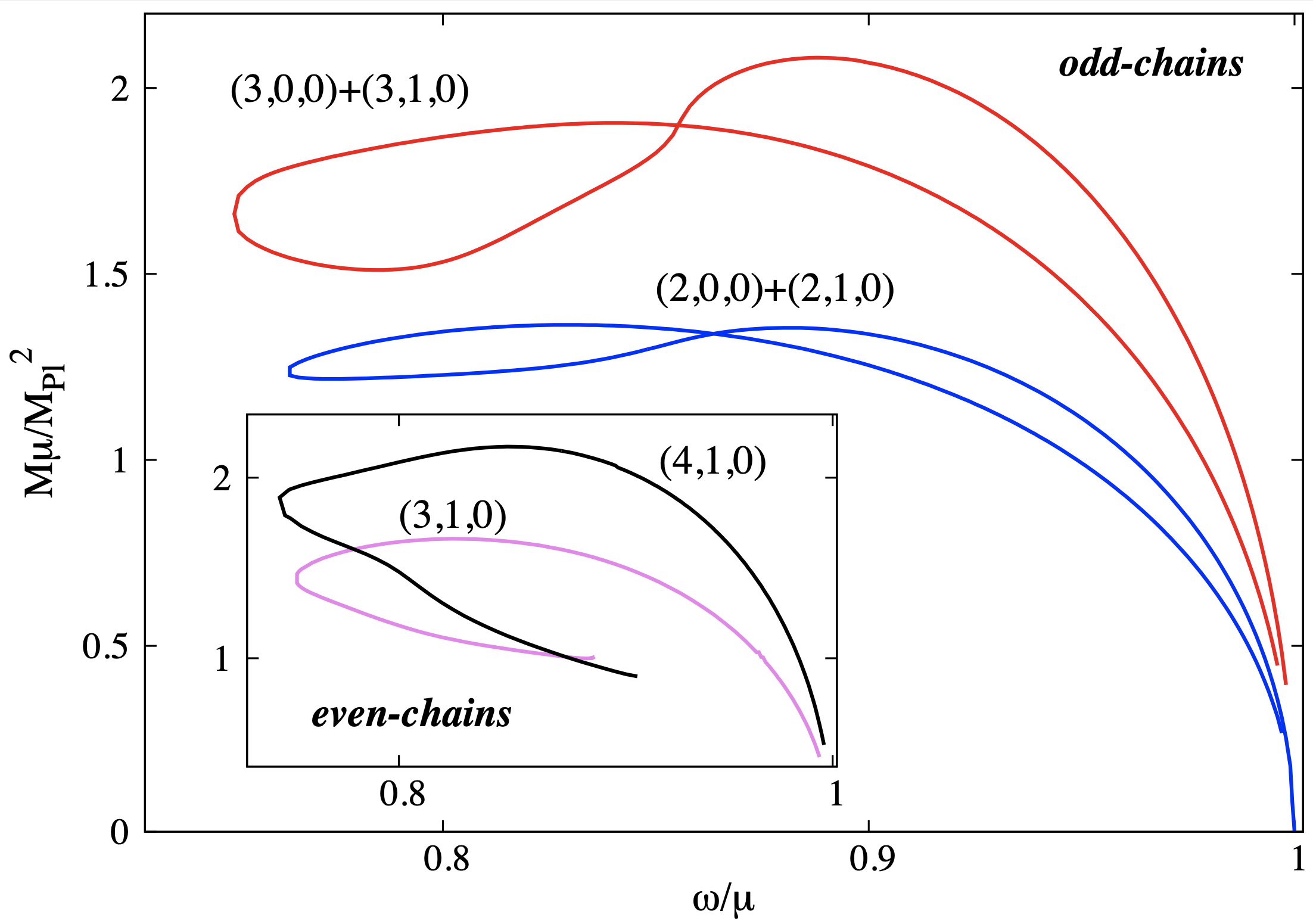}
\caption{\small Domain of existence of the even and odd-chains shown in Fig.~\ref{fig3}. The latter are hybrid multipolar BSs.}
\label{fig4}
\end{center}
\end{figure}

The BSs on the left of Fig.~\ref{fig3} (delimited by the blue dashed line) have nodes along the half $z$-axis.
They are $\Phi_{(N,\ell,0)}$  with $N> \ell+1$. Their domains - see Fig.~\ref{fig4} (inset) - are similar to
those in Fig.~\ref{fig1}. They correspond to the 3$p$ and 4$p$ hydrogen orbitals. From the gravitational side,
these are (even) \textit{chains} of bosonic lumps (see
$e.g.$~\cite{Kleihaus:2003nj,Kleihaus:2004fh,Krusch:2004uf,Shnir:2015aba} for other solitonic chains).
%

{\bf Hybrid Multipolar BSs.}
In QM hybrid orbitals are superpositions of stationary states with different quantum numbers but the same energy; such superposition still yields a stationary state. In hydrogen,  the energy spectrum depends solely on $N$  and hybrid orbitals are of the type $\Psi_{(N,\ell,m)}+\Psi_{(N,\ell',m')}$. Remarkably,  in spite of the non-linearity of the model, multipolar BSs can also yield hybrid configurations, possibly as a consequence of the extended range of frequencies (rather than a single one) of each $\Phi_{(N,\ell,m)}$.

Examples of hybrid multipolar BSs are exhibited on the right of Fig.~\ref{fig3} (delimited by the blue dashed line), corresponding to chains with an odd number of components. They can be interpreted as the superposition of radially excited $Ns$ and $Np$ orbitals. This superposition of states endows the domain of existence of the hybrid solutions with a different structure - Fig.~\ref{fig4} (main panel).
 Instead of the paradigmatic spiraling curve one finds a lace with two ends approaching the maximal frequency, each branch corresponding to the dominance of either of the two states, and with a self-crossing. In particular, we have verified that along one of the branches a bifurcation with the corresponding excited spherical BSs is encountered.

%
%

{\bf Discussion.}
For all multipolar BSs studied, $Q$ is strictly positive. Thus, they are not BS-anti BS configurations.
The Noether charge, moreover, obeys $Q>M$,
for some range of frequencies, $\omega_c<\omega<\mu$, where $\omega_c>\omega_{\rm min}^{(N,\ell,m)}$. Thus, some of the multipolar BSs are stable against fragmentation.

The simplest multipolar BS is the dipolar $\Phi_{(2,1,0)}$. This configuration can be considered a pair of oscillating bosonic lumps with opposite phases. The phase difference yields a repulsive force between them \cite{Bowcock:2008dn}, balancing the gravitational attraction.
The full dynamical status of this and all other multipolar BSs, however, is an open question.

Analogous multipolar configurations are expected in other models, including  flat spacetime $Q$-balls~\cite{Coleman:1985ki} (see also~\cite{Boulle}), Einstein-Klein-Gordon models with self-interactions~\cite{Colpi:1986ye}, gauged BSs~\cite{Jetzer:1989av}, BSs in $AdS$ spacetimes~\cite{Astefanesei:2003qy} and solitons in Einstein-Proca~\cite{Brito:2015pxa} or Einstein-Dirac models~\cite{Finster:1998ws}.


{\bf Acknowledgements.}
This work is supported by the Center for Research and Development in Mathematics and Applications (CIDMA) through the Portuguese Foundation for Science and Technology (FCT - Funda\c c\~ao para a Ci\^encia e a Tecnologia), references UIDB/04106/2020 and, UIDP/04106/2020, and by national funds (OE), through FCT, I.P., in the scope of the framework contract foreseen in the numbers 4, 5 and 6 of the article 23, of the Decree-Law 57/2016, of August 29, changed by Law 57/2017, of July 19.  We acknowledge support  from  the  projects  PTDC/FIS-OUT/28407/2017  and  CERN/FIS-PAR/0027/2019.   This work has further been supported by the European Union’s Horizon 2020 research and innovation (RISE) programme H2020-MSCA-RISE-2017 Grant No. FunFiCO-777740.  The authors would like to acknowledge networking support by the COST Action CA16104.
Ya.S. gratefully acknowledges support by Ministry of Science and High Education of
Russian Federation, project FEWF-2020-0003. JK gratefully acknowledges support by the DFG funded Research Training Group 1620 ``Models of Gravity''. Computations were performed
on the clusters HybriLIT (Dubna) and Blafis (Aveiro).

\bibliography{letterbhlr}



\end{document}